\documentclass[12pt,aps,prd,showpacs,amsmath,amssymb]{revtex4}
\input epsf
\usepackage{graphicx}
\allowdisplaybreaks
\begin{document}
\title{$D_{sJ}(2860)$ From The Semileptonic Decays Of $B_s$ Mesons}
\author{Long-Fei Gan}
 \email{lfgan@nudt.edu.cn}
\author{Jian-Rong Zhang}
\author{Ming-Qiu Huang}
\author{Hong-Bin Zhuo}
\author{Yan-Yun Ma}
\author{Qing-Jun Zhu}
\author{Jian-Xun Liu}
\author{Guo-Bo Zhang}

\affiliation{College of Science, National University of Defense Technology, Changsha, Hunan 410073, People's Republic of China}
\date{\today}
\begin{abstract}
In the framework of heavy quark effective theory, the leading order Isgur-Wise form factors relevant to semileptonic decays of the ground state $\bar{b}s$ meson $B_{s}$ into orbitally excited $D$-wave $\bar{c}s$ mesons, including the newly observed narrow $D^{*}_{s1}(2860)$ and $D^{*}_{s3}(2860)$ states by the LHCb Collaboration, are calculated with the QCD sum rule method. With these universal form factors, the decay rates and branching ratios are estimated.  We find that the decay widths are $\Gamma(B_s\rightarrow D^{*}_{s1}\ell\overline{\nu}) =1.25^{+0.80}_{-0.60}\times10^{-19} \mbox{GeV}$, $\Gamma(B_s\rightarrow D^{'}_{s2}\ell\overline{\nu}) =1.49^{+0.97}_{-0.73}\times10^{-19} \mbox{GeV}$, $\Gamma(B_s\rightarrow D_{s2}\ell\overline{\nu}) =4.48^{+1.05}_{-0.94}\times10^{-17} \mbox{GeV}$, and $\Gamma(B_s\rightarrow D^{*}_{s3}\ell\overline{\nu}) = 1.52^{+0.35}_{-0.31}\times10^{-16} \mbox{GeV}$. The corresponding branching ratios are $\mathcal {B}(B_s\rightarrow D^{*}_{s1}\ell\overline{\nu}) =2.85^{+1.82}_{-1.36}\times 10^{-7}$, $\mathcal {B}(B_s\rightarrow D^{'}_{s2}\ell\overline{\nu}) =3.40^{+2.21}_{-1.66}\times 10^{-7}$, $\mathcal {B}(B_{s}\rightarrow D_{s2}\ell\overline{\nu}) =1.02^{+0.24}_{-0.21}\times 10^{-4}$, and $\mathcal {B}(B_s\rightarrow D^{*}_{s3}\ell\overline{\nu}) = 3.46^{+0.80}_{-0.70}\times 10^{-4}$. The decay widths and branching ratios of corresponding $B^{*}_{s}$ semileptonic processes are also predicted.
\end{abstract}
 \maketitle

\section{Introduction}\label{sec1}
Recently, the LHCb Collaboration released an observation result of two $D_{sJ}(2860)$ resonance states in the process of $B^{0}_{s}\rightarrow \bar{D}^{0}K^{-}\pi^{+}$. They have been considered as mixtures of the $1^-$ and $3^-$ states with the resonance parameters \cite{LHCb141,LHCb142}:
\begin{eqnarray}
m_{D^{*}_{s1}(2860)} &=& (2859\pm12\pm6\pm23) \text{MeV}, \nonumber \\
\Gamma_{D^{*}_{s1}(2860)} &=& (159\pm23\pm27\pm72) \text{MeV}, \nonumber \\
m_{D^{*}_{s3}(2860)} &=& (2860.5\pm2.6\pm2.5\pm6.0) \text{MeV}, \nonumber \\
\Gamma_{D^{*}_{s3}(2860)} &=& (53\pm7\pm4\pm6) \text{MeV}.   \nonumber
\end{eqnarray}
The LHCb Collaboration also announced that this was the first observation of a heavy flavored spin-3 resonance and the first time that any spin-3 particle had been seen to be produced in $B$ decays \cite{LHCb141}. Although $D_{sJ}(2860)$ had been reported before by the BaBar Collaboration \cite{BABAR06,BABAR09}, it has inspired a lot of new interest in studying the spectroscopy of $\bar{c}s$ mesons and the relevant processes \cite{SCLM14,ZCCG14,Wang14,GM14,KZL14}.

Experimentally, copious samples of charm-strange mesons are available from decays of $B^{0}_{_{s}}$ mesons produced at high energy hadron colliders. These have been exploited to study the properties of the orbitally excited $\bar{c}s$ mesons, such as $D_{s1}(2536)^{-}$ and $D^{*}_{s2}(2573)^{-}$ states,  produced in semileptonic decays of $B^{0}_{_{s}}$ mesons \cite{D009}. The results are important not only from the point of view of spectroscopy, but also as they will provide input to future studies of $CP$ violation in the $B^{0}_{s}\rightarrow \bar{D}^{0}K^{-}\pi^{+}$ channel \cite{LHCb142}. Actually, the $b\rightarrow c$ semileptonic processes are the important sources for the determination of the parameters of the standard model, such as Cabibbo-Kobayashi-Maskawa matrix element $|V_{cb}|$. They also provide valuable insight in quark dynamics in the nonperturbative domain of QCD. Just because of these reasons, the semileptonic decays of $B$ and $B_{s}$ mesons have been under investigation for many years \cite{ASS14,Neu92,LLSW97,Dea98,EFG99,Hua99,Col00,Hua04,AAO06,Gan09}.

In this paper, we assume that the newly observed $D^{*}_{s1}(2860)$ and $D^{*}_{s3}(2860)$ mesons are the $1^{-}$ and $3^{-}$ states which are members of the $1D$ family.  Then we use the QCD sum rule method \cite{Shi79} in the framework of heavy quark effective theory (HQET) \cite{LLSW97,Neu94} to study the semileptonic decays of ground  $\bar{b}s$ meson doublet $H(0^{-}, 1^{-})$  into the orbitally $D$-wave excited $\bar{c}s$ meson doublets $F(1^-, 2^-)$ and $X(2^-, 3^-)$ containing one heavy anti-quark and one strange quark. The QCD sum rule approach, incorporation with HQET has been proved to be a successful method which was widely applied to investigate the properties and dynamical processes of heavy hadrons containing a single heavy quark \cite{Neu92}. We shall follow the procedure used in Refs. \cite{Hua04,Gan09,Gan10}, and study the semileptonic decays mentioned above.

The remainder of this paper is organized as follows. After an introduction, we derive the formulae of the weak current matrix elements at the leading order of  HQET in Sec. \ref{sec2}. Then we deduce the three-point sum rules for the relevant universal form factors in Sec. \ref{sec3}. In Sec. \ref{sec4}, we give the numerical results and discussions. The decay rates and branching ratios are also estimated in the final section.

\section{Analytic formulations for semileptonic decay amplitudes $B^{(*)}_{s}\rightarrow (D_{s1}^{*}, D'_{s2})\ell\overline{\nu}$ and $B^{(*)}_{s}\rightarrow (D_{s2}, D_{s3}^{*})\ell\overline{\nu}$} \label{sec2}
The semileptonic decay rate of a $B_{s}$ meson transition into a $D_{s}$ meson is determined by the corresponding matrix elements of the weak vector and axial-vector currents ($V^{\mu} = \overline{c}\gamma^{\mu}b$ and $A^{\mu} = \overline{c}\gamma^{\mu}\gamma_{5}b$) between them. These hadronic matrix elements can be parametrized in terms of some weak form factors. In HQET, the classification of these form factors has been simplified greatly. At the leading order of the heavy quark expansion, the matrix elements involved in the transitions between the $H$ doublet of the $\bar{b}s$ mesons and the $F$ or $X$ doublet of $\bar{c}s$ mesons can be parametrized in terms of only one Isgur-Wise function.

According to the formalism given in Ref. \cite{Fal92}, the heavy-light meson doublets can be expressed as effective operators. For the processes $(B_{s}, B^{*}_{s})\rightarrow (D_{s1}^{*}, D'_{s2})\ell\overline{\nu}$, two heavy-light meson doublets $H$ and $F$ are involved. The operators $P$ and $P^{*}_{\mu}$ that annihilate members of the $H$ doublet with four-velocity $v$ are, in the form,
\begin{equation}\label{operator1}
H_{v} = \frac{1+\rlap/v}{2}[P^*_{\mu}\gamma^{\mu}-P\gamma_5].
\end{equation}
The fields $D^{*}_{1\nu}$ and $D'^{\mu\nu}_{2}$ that annihilate members of the $F$ doublet with four-velocity $v$ are in the representation
\begin{equation}\label{operator2}
F^{\mu}_{v}=\frac{1+\rlap/v}{2}[D'^{\mu\nu}_{2}\gamma_{5}\gamma_{\nu}-D^{*}_{1\nu}\sqrt{\frac{3}{2}}(g^{\mu\nu} -\frac{1}{3}\gamma^{\nu}(\gamma^{\mu}+v^{\mu}))],
\end{equation}
where $\rlap/v = v\cdot\gamma$. For the processes $(B_{s}, B^{*}_{s})\rightarrow (D_{s2}, D_{s3}^{*})\ell \overline{\nu}$, the final heavy hadronic states which annihilated by the operators $D^{\alpha\beta}_{2}$ and $D^{*\mu\nu\sigma}_{3}$ are in another doublet $X$ with four-velocity $v$, namely
\begin{equation}\label{operator3}
 X^{\mu\nu}_{v}=\frac{1+\rlap/v}{2}[D^{*\mu\nu\sigma}_{3}\gamma_{\sigma}-\sqrt{\frac{3}{5}}\gamma_{5}
 D^{\alpha\beta}_{2}(g^{\mu}_{\alpha}g^{\nu}_{\beta}-\frac{\gamma_{\alpha}}{5}g^{\nu}_{\beta}
 (\gamma^{\mu}-v^{\mu})-\frac{\gamma_{\beta}}{5}g^{\mu}_{\alpha}
 (\gamma^{\nu}-v^{\nu}))].
 \end{equation}
At the leading order of heavy quark expansion, the hadronic matrix elements of weak current between states in the doublets $H_{v}$ and $F_{v'}$ can be calculated from
\begin{equation}\label{trace1}
 \bar{h}^{(c)}_{v'}\Gamma h^{(b)}_{v} = \xi(y) \mathrm{Tr}\{v_{\sigma}
 \overline{F}^{(c)\sigma}_{v'}\Gamma H^{(b)}_{v}\},
\end{equation}
while the corresponding matrix elements between states annihilated by fields in $H_{v}$ and $X_{v'}$ are derived from
\begin{equation}\label{trace2}
\bar{h}^{(c)}_{v'}\Gamma h^{(b)}_{v}=\zeta(y)\mathrm{Tr}\{v_{\alpha}v_{\beta}\overline{X}^{(c)\alpha\beta}_{v'} \Gamma H^{(b)}_{v}\},
\end{equation}
where $h^{(Q)}_{v,v'}$ are the heavy quark fields in HQET, and $\overline{X}_{v'}=\gamma_{0}X_{v'}^{\dag}\gamma_{0}$. $v$ is the velocity of the initial meson and $v'$ is the velocity of the final meson in each process. $\Gamma$ denotes the Lorentz structure $\gamma^{\mu}-\gamma^{\mu}\gamma_{5}$  of the weak current. The Isgur-Wise form factors $\xi(y)$ and $\zeta(y)$ are universal functions of the product of velocities $y(=v\cdot v')$. Here we should notice that each side of Eqs. (\ref{trace1}) and (\ref{trace2}) is understood to be inserted between the corresponding initial $\bar{b}s$ and final $\bar{c}s$ states. The hadronic matrix elements of $B_{s}( B^{*}_{s})\rightarrow D^{*}_{s1} (D'_{s2})\ell \overline{\nu}$ can be derived directly from the trace formalism (\ref{trace1}) and are given as
\begin{align}
\label{matrix1}
\frac{\langle D^{*}_{s1}(v^{'},\varepsilon')|(V-A)^{\mu}|B_{s}(v)\rangle}{\sqrt{m_{B_{s}}m_{D^{*}_{s1}}}} = & \frac{1}{3}\sqrt{\frac{3}{2}} \xi(y)\varepsilon^{'*}_{\beta}[v^{\beta } \left((y+2) v'^{\mu }-3 v^{\mu }\right)-\left(y^2-1\right) g^{\beta \mu }\nonumber\\ &-i(y-1) \epsilon ^{\beta \mu \sigma\rho}v_{\sigma} v'_{\rho}],
\\\label{matrix2}
\frac{\langle D^{'}_{s2}(v', \varepsilon')|(V-A)^{\mu}|B_{s}(v)\rangle}{\sqrt{m_{B_{s}}m_{D^{'}_{s2}}}} = & -\xi(y)\varepsilon'_{\alpha\beta}v^{\alpha}\left[(y-1) g^{\beta \mu }-v^{\beta } v'^{\mu }+i \epsilon ^{\beta \mu \sigma\rho}v_{\sigma} v'_{\rho}\right],
\\\label{matrix3}
\frac{\langle D^{*}_{s1}(v',\varepsilon')|(V-A)^{\mu}|B^{*}_{s}(v,\varepsilon)\rangle}{\sqrt{m_{B^{*}_{s}}m_{D^{*}_{s1}}}} = & -\frac{1}{3}\sqrt{\frac{3}{2}} \xi(y)\varepsilon^{'*}_{\beta}\varepsilon_{\sigma}\big[ 3 v^{\beta }v^{\mu } v'^{\sigma }-(y-1) (g^{\beta \sigma } \left(v'^{\mu }+v^{\mu }\right)\nonumber\\ & -g^{\beta \mu } v'^{\sigma }+ 2g^{\mu \sigma } v^{\beta } )-i(v^{\beta } \epsilon ^{\mu \sigma \rho\tau}v_{\rho}v'_{\tau}+2 v^{\mu } \epsilon ^{\beta \sigma \rho\tau}v_{\rho}v'_{\tau} \nonumber\\ & +(y+1) \epsilon ^{\beta \mu \sigma \rho}(v'_{\rho}-v_{\rho}))\big],
\\\label{matrix4}
\frac{\langle D^{'}_{s2}(v',\varepsilon')|(V-A)^{\mu}|B^{*}_{s}(v,\varepsilon)\rangle}{\sqrt{m_{B^{*}_{s}}m_{D^{'}_{s2}}}} = & -\xi(y)\varepsilon^{'*} _{\alpha\beta} \varepsilon_{\sigma} v^{\alpha} [g^{\beta \sigma } \left(v'^{\mu }-v^{\mu }\right)-g^{\beta \mu } v'^{\sigma }+v^{\beta } g^{\mu \sigma } \nonumber\\ & +i \epsilon ^{\beta \mu \sigma \rho}\left( v_{\rho}- v'_{\rho}\right)].
\end{align}
The hadronic matrix elements of $B_{s}(B^{*}_{s})\rightarrow D_{s2}(D_{s3}^{*})\ell \overline{\nu}$ are calculated similarly from Eq. (\ref{trace2}) as follows:
\begin{align}\label{matrix5}
\frac{\langle D_{s2}(v',\varepsilon')|(V-A)^{\mu}|B_{s}(v)\rangle}{\sqrt{m_{B_{s}}m_{D_{s2}}}} = & - \frac{1}{5} \sqrt{\frac{3}{5}} \zeta(y) \varepsilon^{'*}_{\alpha\beta} [\left(y^2-1\right) (v^{\beta }g^{\mu \alpha }+v^{\alpha } g^{\mu \beta })+v^{\alpha }v^{\beta } ((3\nonumber\\ &-2 y) v'^{\mu }+5 v^{\mu })-i (y+1) \left(v^{\alpha } \epsilon ^{\mu \beta \sigma\rho}+v^{\beta } \epsilon ^{\mu \alpha \sigma\rho}\right)v_{\sigma}v'_{\rho}],
\\\label{matrix6}
\frac{\langle D_{s3}^{*}(v',\varepsilon')|(V-A)^{\mu}|B_{s}(v)\rangle}{\sqrt{m_{B_{s}}m_{D_{s3}^{*}}}} = & \zeta(y)\varepsilon^{'*}_{\alpha\beta\rho}v^{\alpha}v^{\beta}[(y+1) g^{\mu \rho }-v^{\rho } v'^{\mu }-i \epsilon ^{\mu \rho \sigma\tau} v_{\sigma}v'_{\tau}],
\\\label{matrix7}
\frac{\langle D_{s2}(v',\varepsilon')|(V-A)^{\mu}|B^{*}_{s}(v,\varepsilon)\rangle}{\sqrt{m_{B^{*}_{s}}m_{D_{s2}}}} = & -\frac{1}{5}\sqrt{\frac{3}{5}} \zeta(y) \varepsilon^{'*}_{\alpha \beta }\varepsilon_{\sigma} [(y+1)(-g^{\mu \beta }v^{\alpha } v'^{\sigma }-v^{\beta } (g^{\mu \alpha } v'^{\sigma }\nonumber\\ & +3 v^{\alpha } g^{\mu \sigma })+(v'^{\mu }-v^{\mu }) \left(v^{\alpha } g^{\sigma \beta }+v^{\beta } g^{\sigma \alpha }\right))+5v^{\mu }v^{\alpha }v^{\beta }  v'^{\sigma }  \nonumber\\ & - i ((2 v^{\alpha } v^{\mu } \epsilon ^{\sigma \beta \rho \tau}-v^{\alpha } v^{\beta } \epsilon ^{\mu \sigma \rho \tau} +2 v^{\beta } v^{\mu } \epsilon ^{\sigma \alpha \rho \tau})v_{\rho}v'_{\tau} \nonumber\\ & +(y-1)( v^{\alpha } \epsilon ^{\mu \sigma \beta \rho}+ v^{\beta } \epsilon ^{\mu \sigma \alpha \rho})(v_{\rho}+v'_{\rho}))],
\\\label{matrix8}
\frac{\langle D^{*}_{s3}(v',\varepsilon')|(V-A)^{\mu}|B^{*}_{s}(v,\varepsilon)\rangle}{\sqrt{m_{B^{*}_{s}}m_{D^{*}_{s3}}}} = & \zeta(y)\varepsilon^{'*}_{\alpha\beta\rho} \varepsilon_{\sigma}v^{\alpha}v^{\beta}[g^{\rho \sigma } (v'^{\mu }+v^{\mu })-g^{\mu \rho } v'^{\sigma }- g^{\mu \sigma }v^{\rho } \nonumber\\ & +i \epsilon ^{\mu \rho \sigma \tau}\left(v'_{\tau}+v_{\tau}\right)].
\end{align}
In these matrix elements, $\varepsilon_{\alpha}$ ($\varepsilon'_{\alpha}$) is the polarization vector of the initial (final) vector meson while $\varepsilon'_{\alpha\beta}$ and $\varepsilon'_{\alpha\beta\rho}$ are the polarization tensors of final tensor mesons. In the derivation of the matrix elements and formulae below, we have used a \emph{Mathematica} package called \emph{FeynCalc} \cite{FeynC}.  The only unknown factors in the matrix elements above are the Isgur-Wise form factors $\xi(y)$ and $\zeta(y)$ which should be determined through nonperturbative methods. In the following section, we will employ the QCD sum rule approach to estimate them.

It is worth noting that the matrix elements of the weak current between $B_s$ mesons and
excited $D_s$ mesons vanish at zero recoil in the heavy quark limit due to the heavy quark symmetry. The heavy quark $1/m_Q$ corrections, which can be finite at this kinematic point, may provide significant modification of the decay rates calculated in the heavy quark limit. Meanwhile, one could expect the calculations of $1/m_Q$ corrections especially for so many decay processes considered in this work are tedious as one has to deal with lots of sub-leading order form factors and they all should be estimated by nonperturbative methods. On the other hand, it can be expected that the $1/m_Q$ corrections might still be under control seeing from some previous works, e.g. \cite{LLSW97,Hua04}. Hence the calculations in this work have been confined at the leading order of the heavy quark expansion.

\section{Form factors from HQET sum rules}\label{sec3}
In order to apply QCD sum rules to study the heavy mesons, we must choose appropriate interpolating currents to represent them. Here we adopt the interpolating currents proposed in Ref. \cite{Dai97} based on the study of Bethe-Salpeter equations for heavy mesons in HQET. Following the remarks given in Ref. \cite{Gan09}, we take the interpolating currents that create heavy mesons in the $H$, $F$ and $X$ doublets as
\begin{align}
\label{current1}
J^{\dag}_{0,-,1/2} &=\frac{1}{\sqrt{2}}\bar{h}_{v}\gamma_{5}s,
\\\label{current2}
J^{\alpha\dag}_{1,-,1/2} &=\frac{1}{\sqrt{2}}\bar{h}_{v}\gamma^{\alpha}_{t}s,
\\\label{current3}
J^{\alpha\dag}_{1,-,3/2} &=-\sqrt{\frac{3}{4}}\bar{h}_{v}(D^{\alpha}_{t}-\frac{1}{3} \gamma^{\alpha}_{t}\!\not\!\!{D}_{t})\!\not\!\!{D}_{t}s,
\\ \label{current4}
J^{\alpha\beta\dag}_{2,-,3/2} &=-\frac{1}{\sqrt{2}}T^{\alpha\beta,\mu\nu}\bar{h}_{v}\gamma_{5} \gamma_{t\mu}D_{t\nu}\!\not\!\!{D}_{t}s,
\\ \label{current5}
J^{\alpha\beta\dag}_{2,-,5/2} &=-\sqrt{\frac{5}{6}}T^{\alpha\beta,\mu\nu}\overline{h}_{v}\gamma_{5}
(D_{t\mu}D_{t\nu}-\frac{2}{5}D_{t\mu}\gamma_{t\nu}\rlap/D_{t})s ,
\\ \label{current6}
J^{\alpha\beta\lambda\dag}_{3,-,5/2} &=-\frac{1}{\sqrt{2}}T^{\alpha\beta\lambda,\mu\nu\sigma}\overline{h}_{v} \gamma_{t\mu} D_{t\nu}D_{t\sigma}s,
\end{align}
where $D^{\alpha}_{t}=D^{\alpha}-v^{\alpha}(v\cdot D)$ is the transverse component of the covariant derivative with respect to the velocity of the meson. The tensors $T^{\alpha\beta,\mu\nu}$ and $T^{\alpha\beta\lambda,\mu\nu\sigma}$ are used to symmetrize the indices and given by
\begin{eqnarray}
\label{tensor1}
T^{\alpha\beta,\mu\nu}&=&\frac{1}{2}(g^{\alpha\mu}_{t}g^{\beta\nu}_{t}
+g^{\alpha\nu}_{t}g^{\beta\mu}_{t})-\frac{1}{3}g^{\alpha\beta}_{t}g^{\mu\nu}_{t},\\
\label{tensor2}
T^{\alpha\beta\lambda,\mu\nu\sigma}&=&\frac{1}{6}(g^{\alpha\mu}_{t}g^{\beta\nu}_{t}g^{\lambda\sigma}_{t} +g^{\alpha\mu}_{t}g^{\beta\sigma}_{t}g^{\lambda\nu}_{t}+g^{\alpha\nu}_{t}g^{\beta\mu}_{t}g^{\lambda\sigma}_{t} +g^{\alpha\nu}_{t}g^{\beta\sigma}_{t}g^{\lambda\mu}_{t}+g^{\alpha\sigma}_{t}g^{\beta\nu}_{t}g^{\lambda\mu}_{t}+
g^{\alpha\sigma}_{t}g^{\beta\mu}_{t}g^{\lambda\nu}_{t}) \nonumber\\ & & -\frac{1}{15}(g^{\alpha\beta}_{t} g^{\mu\nu}_{t}g^{\lambda\sigma}_{t}+g^{\alpha\beta}_{t} g^{\mu\sigma}_{t} g^{\lambda\nu}_{t}+
g^{\alpha\beta}_{t}g^{\nu\sigma}_{t}g^{\lambda\mu}_{t}+g^{\alpha\lambda}_{t}g^{\mu\nu}_{t}g^{\beta\sigma}_{t}
+g^{\alpha\lambda}_{t}g^{\mu\sigma}_{t}g^{\beta\nu}_{t}\nonumber\\& &+g^{\alpha\lambda}_{t}g^{\nu\sigma}_{t} g^{\beta\mu}_{t}+g^{\beta\lambda}_{t}g^{\mu\nu}_{t}g^{\alpha\sigma}_{t}+g^{\beta\lambda}_{t} g^{\mu\sigma}_{t} g^{\alpha\nu}_{t}+g^{\beta\lambda}_{t}g^{\nu\sigma}_{t}g^{\alpha\mu}_{t}),
\end{eqnarray}
where $g^{\alpha\beta}_{t} = g^{\alpha\beta} - v^{\alpha} v^{\beta}$ is the transverse part of the metric tensor relative to the velocity of the heavy meson.

These currents have non-vanishing projections only to the corresponding states of the HQET in the $m_{Q}\rightarrow\infty$ limit, without mixing with states of the same quantum number but different $s_{l}$ \cite{Dai97}. Thus we can define one-particle-current couplings as follows:
\begin{align}
\label{const1}
\langle H_{s0}(v,\varepsilon)|J^{\dag}_{0,-,1/2}|0\rangle &= f_{0,-,1/2}\sqrt{m_{H_{s0}}} , &\text{ for}\; J^{P}=0^{-};
\\ \label{const2}
\langle H_{s1}(v,\varepsilon)|J^{\alpha\dag}_{1,-,1/2}|0\rangle &= f_{1,-,1/2}\sqrt{m_{H_{s1}}} \varepsilon^{*\alpha}, &\text{ for}\; J^{P}=1^{-};
\\ \label{const3}
\langle H^{*}_{s1}(v,\varepsilon)|J^{\alpha\dag}_{1,-,3/2}|0\rangle &= f_{1,-,3/2}\sqrt{m_{H^{*}_{s1}}} \epsilon^{*\alpha}, &\text{ for}\; J^{P}=1^{-};
\\ \label{const4}
\langle H'_{s2}(v,\varepsilon)|J^{\alpha\beta\dag}_{2,-,3/2}|0\rangle &= f_{2,-,3/2} \sqrt{m_{H'_{s2}}} \epsilon^{*\alpha\beta}, &\text{ for}\; J^{P}=2^{-};
\\ \label{const5}
\langle H_{s2}(v,\varepsilon)|J^{\alpha\beta\dag}_{2,-,5/2}|0\rangle &= f_{2,-,5/2}\sqrt{m_{H_{s2}}} \epsilon^{*\alpha\beta}, &\text{ for}\; J^{P}=2^{-};
\\ \label{const6}
\langle H^{*}_{s3}(v,\varepsilon)|J^{\alpha\beta\lambda\dag}_{3,-,5/2}|0\rangle &= f_{3,-,5/2}\sqrt{m_{H^{*}_{s3}}} \epsilon^{*\alpha\beta\lambda}, &\text{ for}\; J^{P}=3^{-}.
\end{align}
The decay constants $f_{0,-,1/2}$, $f_{1,-,1/2}$, $f_{1,-,3/2}$, $f_{2,-,3/2}$, $f_{2,-,5/2}$, and $f_{3,-,5/2}$ are low-energy parameters which are determined by the dynamics of the light degree of freedom.

With these currents, we can now estimate the Isgur-Wise functions $\xi(y)$ and $\zeta(y)$ from QCD sum rules. First comes the $\xi(y)$. The jumping-off point is the following three-point correlation function:
\begin{align}\label{correlator1}
\Xi^{\alpha\mu}(\omega,\omega^{'},y) & =i^{2}\int d^{4}xd^{4}ze^{i(k^{'}\cdot x-k\cdot z)}\langle0|T[J^{\alpha}_{1,-,3/2}(x) J^{\mu(v,v^{'})}_{V,A}(0)J^{\dag}_{0,-,1/2}(z)]|0\rangle\nonumber\\ & = \Xi_{1}(\omega,\omega^{'},y)\mathcal {L}^{\alpha\mu}_{\xi(V,A)},
\end{align}
where $J^{\mu(v,v^{'})}_{V}=h(v^{'})\gamma^{\mu}h(v)$ and $J^{\mu(v,v^{'})}_{A}=h(v^{'})\gamma^{\mu}\gamma_{5}h(v)$ are the weak currents. $J_{0,-,1/2}$ and $J^{\alpha}_{1,-,3/2}$ are the interpolating currents defined in Eqs. (\ref{current1}) and (\ref{current3}). Here it is worth noting that $\xi(y)$ can also be estimated by choosing the interpolating current (\ref{current2}) for the initial state and  the current (\ref{current4}) for the final state because of the heavy quark symmetry. $\Xi_{1}(\omega,\omega^{'},y)$ is an analytic function in $\omega=2v \cdot k$ and $\omega'=2v' \cdot k'$, and is not continual when $\omega$ and $\omega'$ locate on the positive real axis. $k$($=P-m_{b}v$) and $k^{'}$($=P'-m_{c}v'$) are the residual momenta of the initial and final meson states, respectively. The scalar function $\Xi_{1}(\omega,\omega^{'},y)$ also depends on the velocity transfer $y=v \cdot v'$. $\mathcal {L}^{\alpha\mu}_{\xi(V,A)}$ are the Lorentz structures.

To calculate the phenomenological or physical part of the correlator (\ref{correlator1}), we insert two complete sets of intermediate states with the same quantum numbers as the currents $J_{0,-,1/2}$ and $J^{\alpha}_{1,-,3/2}$, then isolate the contribution from the double pole at $\omega=2 \bar{\Lambda}_{-,1/2}$, $\omega'=2 \bar{\Lambda}_{-,3/2}$:
\begin{align}\label{pheno}
\Xi^{\alpha\mu}(\omega,\omega^{'},y) & = \frac{f_{0,-,1/2}f_{1,-,3/2}}{(2\bar{\Lambda}_{-,1/2}-\omega-i \epsilon)(2\bar{\Lambda}_{-,3/2}-\omega'-i \epsilon)}\xi(y)\mathcal{L}^{\alpha\mu}_{\xi}+\cdots,
\end{align}
where ``$\cdots$" denotes the contribution from higher resonances and continuum states while $f_{1,-,3/2}$ is the decay constant defined in Eq. (\ref{const1}). As we can see from the Eqs. (\ref{correlator1}) and (\ref{pheno}), the pole contribution to $\Xi_{1}(\omega,\omega^{'},y)$ is proportional to the universal function $\xi(y)$. The QCD sum rule then can be constructed directly from $\Xi_{1}(\omega,\omega^{'},y)$ by isolating the Lorentz structures.

The theoretical side of the correlator is calculated by means of the operator product expansion. The perturbative part can be expressed as a double dispersion integral in $\nu$ and $\nu^{'}$ plus possible subtraction terms. Therefore the theoretical expression for the correlation function in (\ref{correlator1}) is of the form
\begin{equation}\label{theo}
\Xi_{1}^{\text{theo}}(\omega,\omega^{'},y)\simeq\int d\nu d\nu^{'} \frac{\rho^{\text{pert}} (\nu,\nu^{'},y)} {(\nu-\omega-i\varepsilon) (\nu^{'}-\omega^{'}-i\varepsilon)}+ \text{subtractions} + \Xi^{\text{cond}}_{1}(\omega,\omega^{'},y).
\end{equation}
The first term on the right hand side of Eq. (\ref{theo}) is the perturbative contribution while ``subtractions" means the subtraction terms resulted from the dispersion relation. The third term $\Xi^{\text{cond}}_{1}(\omega,\omega^{'},y)$ denotes the contribution from quark and gluon condensations. The perturbative spectral density $\rho^{\text{pert}}(\nu,\nu^{'},y)$ can be calculated straightforwardly from HQET Feynman rules. At the leading order of perturbation and heavy quark expansion, we obtain the perturbative spectral density of the sum rule for $\xi(y)$ as
\begin{align}\label{perturb1}
\rho^{\text{pert}}_{\xi}(\nu,\nu',y) = &\frac{3}{8 \pi ^2} \frac{1}{(y+1)^{3/2}(y-1)^{5/2}} \nu' \left[3 \nu^2+(2 y+1) \left(\nu'\right)^2-2 (2 y \nu +\nu ) \nu '\right] \nonumber \\ & \times\Theta(\nu)\Theta(\nu')\Theta(2y\nu\nu'-\nu^{2}-\nu'^{2}).
\end{align}
Assuming quark-hadron duality, the contribution from higher resonances is usually approximated by the integration of the perturbative spectral density above some threshold. Equating the phenomenological and theoretical representations, the contribution of higher resonances in the phenomenological expression (\ref{pheno}) can be eliminated. Following the arguments in Refs. \cite{Neu92,Blo93}, we can not directly assume local duality between the perturbative and the hadronic spectral densities, but first integrate the spectral density over the ``off-diagonal" variable $\nu_{-}=\nu-\nu^{'}$, keeping the ``diagonal" variable $\nu_{+}=\frac{\nu+\nu^{'}}{2}$ fixed. Then the quark-hadron duality is assumed for the integration of the spectral density in $\nu_{+}$. The integration region is restricted by the $\Theta$ functions above in terms of the variables $\nu_{-}$ and $\nu_{+}$, and usually the triangular region defined by the bounds: $0\leq \nu_{+}\leq \omega_{c}$, $-2\sqrt{\frac{y-1}{y+1}}\nu_{+}\leq \nu_{-}\leq
2\sqrt{\frac{y-1}{y+1}}\nu_{+}$ is chosen. A double Borel transformation in $\omega$ and $\omega^{'}$ is performed on both sides of the sum rule, in which for simplicity we take the Borel parameters equal \cite{Neu92,Hua99,Col00}: $T_{1}=T_{2}=2T$. It eliminates the subtraction terms in the dispersion integral (\ref{theo}) and improves the convergence of the operator product expansion series. Our calculations are confined at the leading order of perturbation. Among the operators in the operator product expansion series, only those with dimension $D \leq  5$ are included. For the condensates of higher dimension ($D > 5$), their values are negligibly small and their contributions are suppressed by the double Borel transformation. So they can be safely omitted. Finally, we obtain the sum rule for the form factor $\xi(y)$ as follows:
\begin{align}
\label{rule1}
\xi(y)f_{0,-,1/2}f_{1,-,3/2}e^{-(\bar{\Lambda}_{0,-,1/2}+\bar{\Lambda}_{1,-,3/2})/T} = & \frac{1}{16\pi^{2}} \frac{1}{(y+1)^{3}} \int^{\omega_{c1}}_{0}d\nu_{+}e^{-\nu_{+}/T}[\nu_{+}^{4} -4m_{s}(y+1)\nonumber \\ &\times\nu_{+}^{3} +3m^{2}_{s}(y+1)\nu_{+}^{2}] +\frac{T}{24}\frac{3y-4}{(y+1)^{2}} \langle \frac{\alpha_{s}}{4\pi} GG\rangle.
\end{align}

The derivation of the sum rule for $\zeta(y)$ is totally similar. Only the correlation function one needs to consider now is
\begin{equation}\label{correlator2}
i^{2}\int d^{4}xd^{4}ze^{i(k^{'}\cdot x-k\cdot z)}\langle0|T[J^{\alpha\beta}_{2,-,5/2}(x) J^{\mu(v,v^{'})}_{V, A}(0) J^{\dag}_{0,-,1/2}(z)|0\rangle=\Xi_{2}(\omega,\omega^{'},y)\mathcal{L}^{\alpha\beta\mu}_{V,A},
\end{equation}
where $J^{\mu(v,v^{'})}_{V, A}$ are also the weak currents. $J_{0,-,1/2}$ and $J^{\alpha\beta}_{2,-,5/2}$ are the interpolating currents defined in Eqs. (\ref{current1}) and (\ref{current5}). By repeating the above procedure, we reach the perturbative spectral density as below:
\begin{align}\label{perturb2}
\rho^{\text{pert.}}_{\zeta}(\nu,\nu',y) = & \frac{3 }{2 \pi ^2} \frac{1}{(y+1)^{7/2}(y-1)^{5/2}} [5 \nu ^3+ \left(2 y^2-2 y+1\right) (3\nu +\nu') \left(\nu '\right)^2 \nonumber \\ & +3 (1-4 y) \nu ^2 \nu ']\Theta(\nu)\Theta(\nu')\Theta(2y\nu\nu'-\nu^{2}-\nu'^{2}).
\end{align}
Then the sum rule for $\zeta(y)$ appears to be
\begin{align}\label{rule2}
\zeta(y)f_{0,-,1/2}f_{2,-,5/2}e^{-(\bar{\Lambda}_{0,-,1/2}+\bar{\Lambda}_{2,-,5/2})/T} = & \frac{1}{8\pi^{2}} \frac{1}{(y+1)^{4}} \int^{\omega_{c2}}_{0}d\nu_{+}e^{-\nu_{+}/T}[3\nu_{+}^{4} +2m_{s}(y+1) \nu_{+}^{3}\nonumber \\ & +6m^{2}_{s}(y+1)\nu_{+}^{2}] +\frac{T}{3\times2^{5}} \frac{13y-25}{(y+1)^{3}} \langle \frac{\alpha_{s}}{4\pi} GG\rangle.
\end{align}

\section{Numerical results and discussions}\label{sec4}
Now comes the evaluation of the sum rules derived in the previous section numerically. First, we specify the input parameters in our calculation. For the vacuum condensation parameters, we adopt the standard values: $\label{qcond}\langle\overline{q}q\rangle=-(0.24)^{3}\,\mbox{GeV}^{3}$, $\label{gcond}\langle\alpha_{s}GG\rangle=0.04\,\mbox{GeV}^{4}$, and $\label{scond}\langle\bar{s}s\rangle=(0.8\pm 0.2)\,\langle\overline{q}q\rangle$. The mass of the strange quark is $m_{s}=150\,\mbox{MeV}$. For masses of the initial $B_{s}$ and $B^{*}_{s}$ mesons, we use $M_{B_{s}}=5366.7 \,\mbox{MeV}$ and $M_{B^{*}_{s}}=5415.4\,\mbox{MeV}$ \cite{PDG14}. For masses of the final $D^{*}_{s1}$, $D^{'}_{s2}$, $D_{s2}$, and $D^{*}_{s3}$ mesons, we use $M_{D^{*}_{s1}}=2859\,\mbox{MeV}$ \cite{LHCb141}, $M_{D^{'}_{s2}}=2810\,\mbox{MeV}$ \cite{ZCCG14}, $M_{D_{s2}}=2820\,\mbox{MeV}$ \cite{ZCCG14}, and $M_{D^{*}_{s3}}=2860.5\,\mbox{MeV}$ \cite{LHCb141}.

In order to obtain information of Isgur-Wise function $\xi(y)$ and $\zeta(y)$ with less systematic uncertainty, we can divide the three-point sum rules (\ref{rule1}) and (\ref{rule2}) with the square roots of relevant two-point sum rules for the decay constants, as many authors did \cite{Neu92,Hua99,Col00}. This can not only reduce the number of input parameters but also improve stabilities of the three-point sum rules. In the calculation of $\xi(y)$, the two-point QCD sum rules we need are
\begin{align}\label{conrule1}
f^{2}_{0,-,1/2}e^{-2\bar{\Lambda}_{0,-,1/2}/T} = & \frac{3}{16\pi^{2}}\int^{\omega_{0}}_{2m_{s}}d\nu e^{-\nu/T} (\nu^{2}+2m_{s}\nu-2m^{2}_{s})-\frac{1}{2}\langle\bar{s}s\rangle(1-\frac{m_{s}}{2T}+\frac{m^{2}_{s}}{2T^{2}})\nonumber\\
& +\frac{m^{2}_{0}}{8T^{2}} \langle\bar{s}s\rangle (1-\frac{m_{s}}{3T}+\frac{m^{2}_{s}}{3T^{2}}) -\frac{m_{s}}{16T^{2}}\langle\frac{\alpha_{s}}{4\pi}GG\rangle (2\gamma_{E}-1-\text{ln}\frac{T^{2}}{\mu^{2}})
\end{align}
in the Ref. \cite{DHLZ03} and
\begin{align}\label{conrule2}
f^{2}_{1,-,3/2}e^{-2\bar{\Lambda}_{1,-,3/2}/T} = & \frac{7}{2560\pi^{2}}\int^{\omega_{1}}_{2m_{s}}d\nu e^{-\nu/T} (\nu^{6} +2m_s\nu^{5} -10m^{2}_{s}\nu^{4}) -\frac{T^{3}}{2} \langle\frac{\alpha_{s}}{4\pi}GG\rangle
\end{align}
in the Ref. \cite{ZCCG14}. Here the cutoff parameter $\mu$ is fixed at $1\text{GeV}$ and the Euler parameter $\gamma_{E} = 0.577$. In order to calculate $\zeta(y)$, we need the two-point QCD sum rules (\ref{conrule1}) and
\begin{align}\label{conrule3}
f^{2}_{2,-,5/2}e^{-2\bar{\Lambda}_{2,-,5/2}/T} = & \frac{1}{640\pi^{2}}\int^{\omega_{2}}_{2m_{s}}d\nu e^{-\nu/T} (\nu^{6} +2m_s\nu^{5} -10m^{2}_{s}\nu^{4}) -\frac{3T^{3}}{8} \langle\frac{\alpha_{s}}{4\pi}GG\rangle
\end{align}
in the Ref. \cite{ZCCG14}.

\begin{figure}
\begin{center}
\begin{tabular}{ccc}
\begin{minipage}{7cm} \epsfxsize=7cm
\centerline{\epsffile{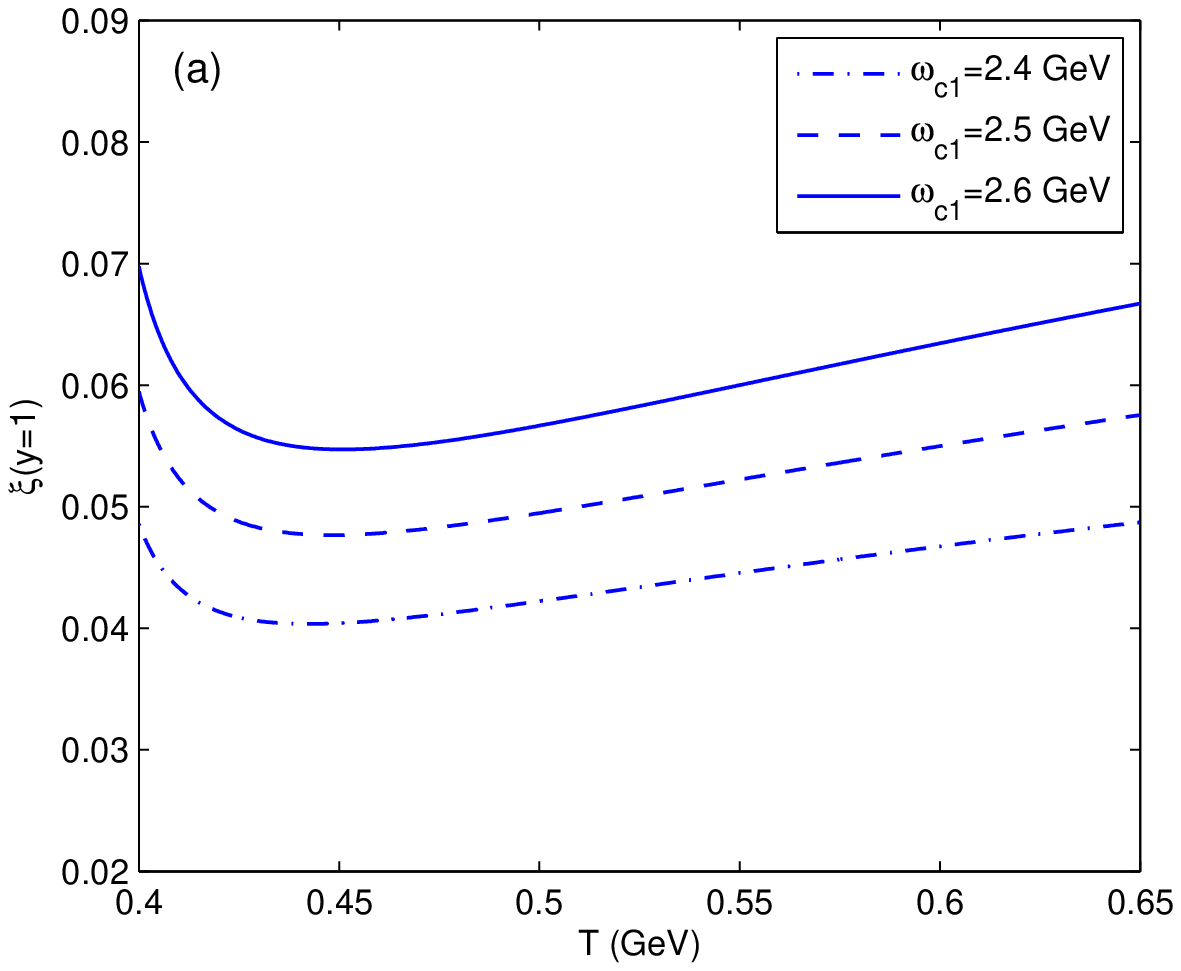}}
\end{minipage}& &
\begin{minipage}{7cm} \epsfxsize=7cm
\centerline{\epsffile{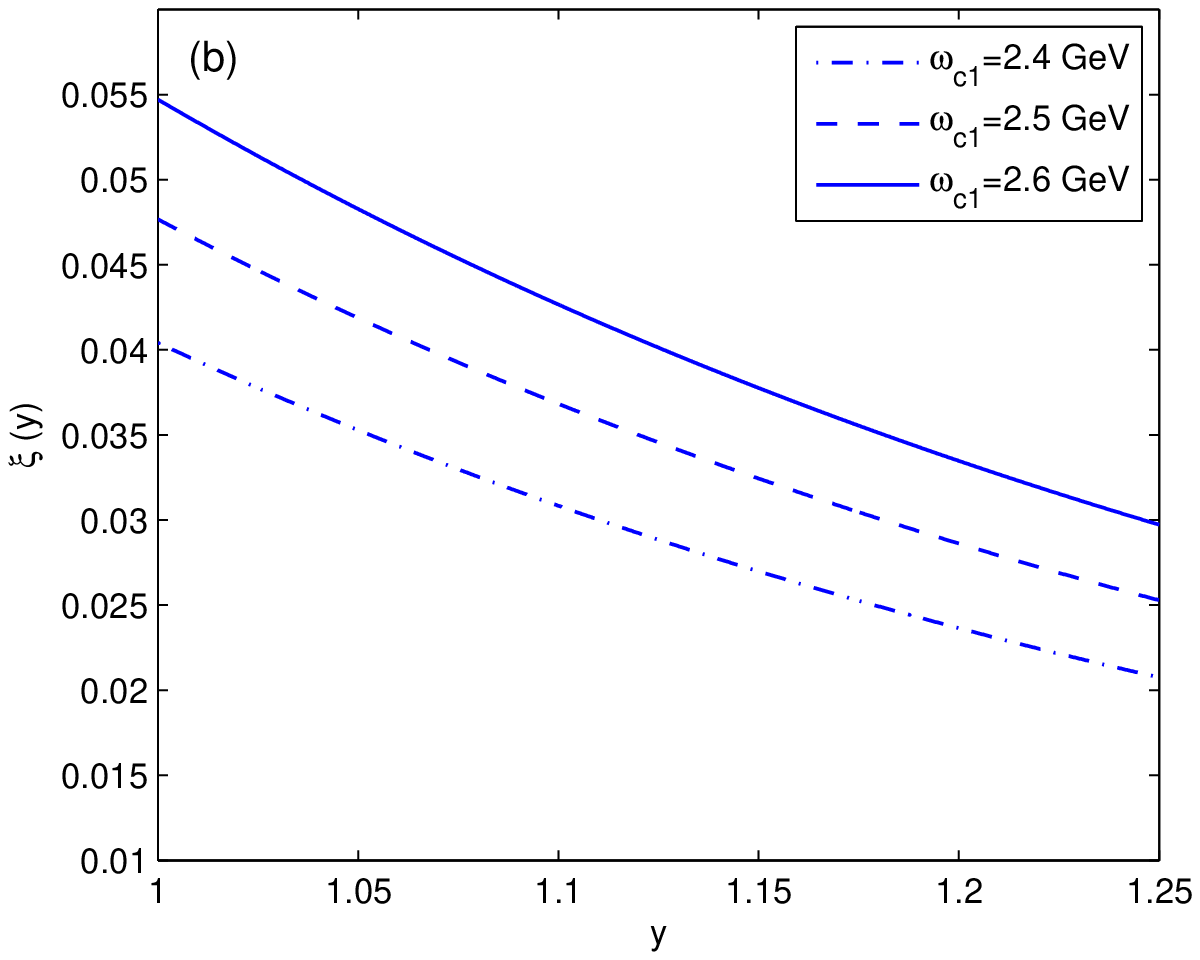}}
\end{minipage}
\end{tabular}
\caption{(a) {\it Dependence of  $\xi(y)$  on Borel parameter $T$ at $y=1$.} (b) {\it Prediction for the Isgur-Wise functions $\xi(y)$ at} $T=0.45\,\mbox{GeV}$.}\label{fig1}
\end{center}
\end{figure}

\begin{figure}
\begin{center}
\begin{tabular}{ccc}
\begin{minipage}{7cm} \epsfxsize=7cm
\centerline{\epsffile{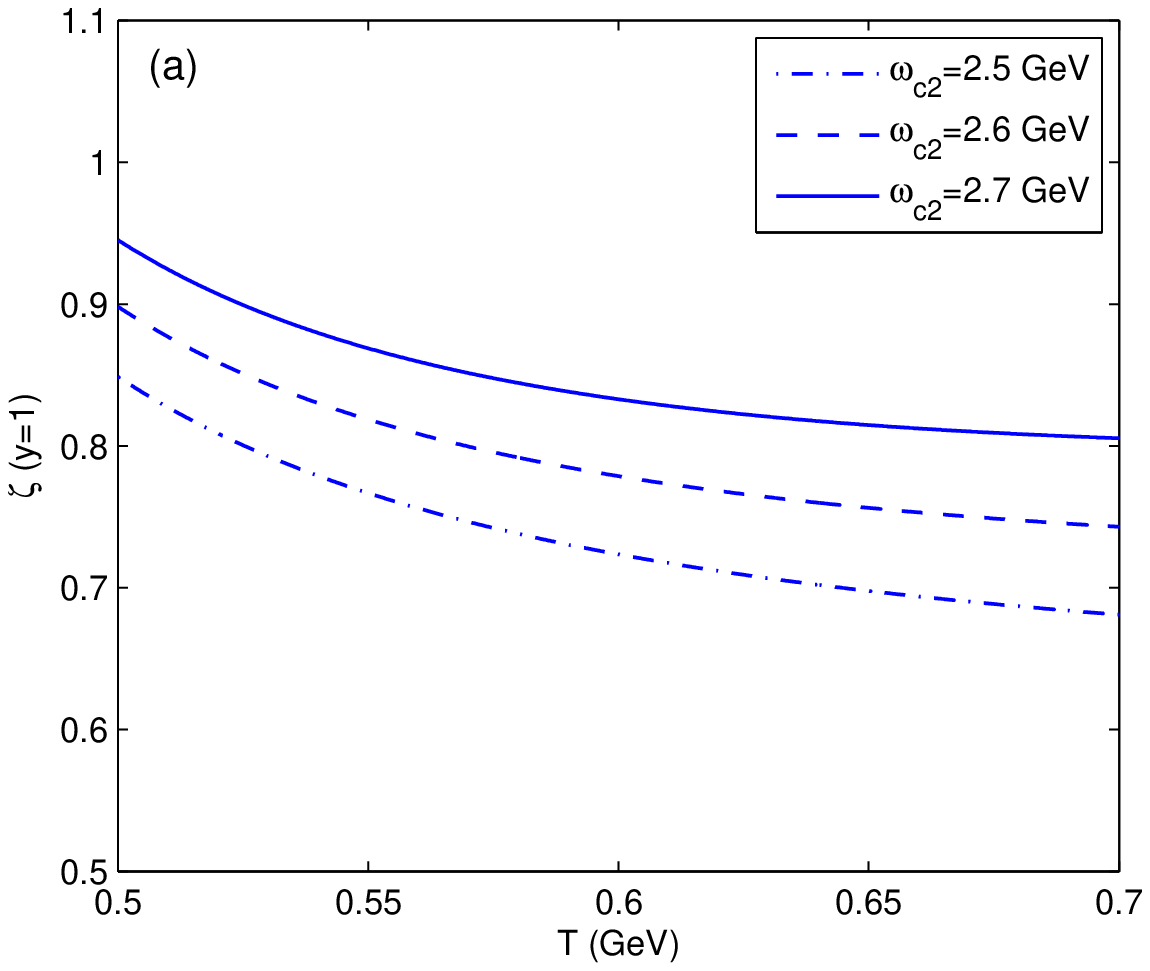}}
\end{minipage}& &
\begin{minipage}{7cm} \epsfxsize=7cm
\centerline{\epsffile{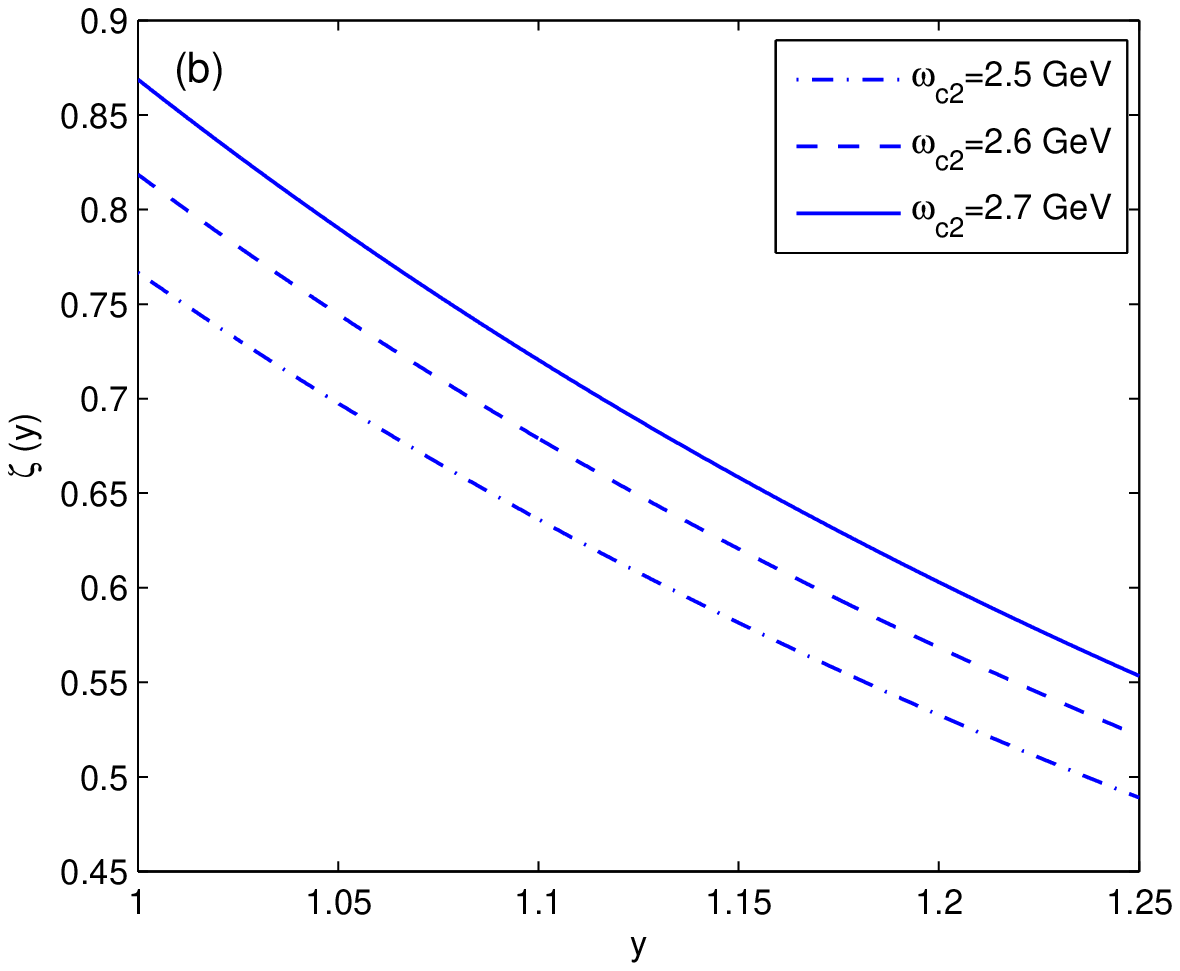}}
\end{minipage}
\end{tabular}
\caption{(a) {\it Dependence of $\zeta(y)$ on Borel parameter $T$ at $y=1$.} (b) {\it Prediction for the Isgur-Wise functions $\zeta(y)$ at} $T=0.55\,\mbox{GeV}$.}\label{fig2}
\end{center}
\end{figure}

After the divisions have been done, the Isgur-Wise functions $\xi(y)$ and $\zeta(y)$ depend only on the Borel parameter $T$ and the continuum thresholds. The determination of the Borel parameter is an important step of the QCD sum rule method. After a careful analysis, we find that the sum rule for $\xi(y)$ works well in a sum rule ``window": $0.4\,\mbox{GeV} < T <0.6\,\mbox{GeV}$, which overlaps with that of the two-point sum rule (\ref{conrule1}) \cite{DHLZ03}. For the sum rule of $\zeta(y)$, we choose the ``window" as $0.5\,\mbox{GeV} < T < 0.7\,\mbox{GeV}$. Note that the Borel parameters in the three-point sum rules are twice of those in the two-point sum rules. In the evaluation, we have taken $2.0 \, \mbox{GeV} < \omega_{0} < 2.4 \, \mbox{GeV}$, $2.8 \, \mbox{GeV} < \omega_{1} < 3.2 \, \mbox{GeV}$, and $3.2 \, \mbox{GeV} < \omega_{2} < 3.6 \, \mbox{GeV}$ \cite{Gan09}. The regions of these continuum thresholds are fixed by analyzing the corresponding two-point sum rules \cite{DHLZ03}. Following the discussions in Refs. \cite{Blo93,Neu92}, the upper limit $\omega_{c1}$ for $\nu_{+}$ in Eq. (\ref{rule1}) and $\omega_{c2}$ in Eq. (\ref{rule2}) should be evaluated in the regions $\frac{1}{2}[(y+1)-\sqrt{y^{2}-1}]\omega_{0}\leqslant\omega_{c1}\leqslant\frac{1}{2} (\omega_{0} +\omega_{1})$ and $\frac{1}{2}[(y+1)-\sqrt{y^{2}-1}]\omega_{0}\leqslant\omega_{c2}\leqslant\frac{1}{2} (\omega_{0} +\omega_{2})$. So they can be fixed in the regions $2.4\,\mbox{GeV}<\omega_{c1}<2.6\,\mbox{GeV}$ and $2.5\,\mbox{GeV}<\omega_{c2}<2.7\,\mbox{GeV}$ . Taking account of all these parameters, we get the results that are shown in Fig. \ref{fig1} and Fig. \ref{fig2}, where we have fixed $\omega_{0}=2.1\,\mbox{GeV}$ in the two-point sum rule (\ref{conrule1}), $\omega_{1}=3.0\,\mbox{GeV}$ in Eq. (\ref{conrule2}), and $\omega_{2}=3.4\,\mbox{GeV}$ in Eq. (\ref{conrule3}).

The curves for $\xi(y)$ and $\zeta(y)$ shown in the figures above can be parametrized by the linear approximations:
\begin{align}
\label{linear1}
\xi(y)&=\xi(1)-\rho^{2}_{\xi}(y-1),\text{ }
\xi(1)=0.046\pm0.009,\text{ }\rho^{2}_{\xi}=0.089,
\\ \label{linear2}
\zeta(y)&=\zeta(1)-\rho^{2}_{\zeta}(y-1),\text{ }
\zeta(1)=0.803\pm0.067,\text{ }\rho^{2}_{\zeta}=1.18.
\end{align}
The errors are resulted from the sum rule working ``window" and reflect the uncertainty due to the continuum threshold $\omega_{c}$ and the Borel parameter $T$. The uncertainty due to the variation of the QCD and HQET parameters is not included here, which may reach $5\%$ or more \cite{Hua04}. Using the linear approximations for the universal form factors above, one can calculate the semileptonic decay rates of processes $B_{s}(B^{*}_{s})\rightarrow D^{*}_{s1}(D^{'}_{s2})\ell\overline{\nu}$ and $B_{s}(B^{*}_{s})\rightarrow D_{s2}(D^{*}_{s3})\ell\overline{\nu}$. For this purpose, we have to derive firstly the formulae for the differential decay rates of these processes in terms of the Isgur-Wise functions $\xi(y)$  and $\zeta(y)$ from the matrix elements (\ref{matrix1})-(\ref{matrix8}) given in Sec. \ref{sec2}. After some derivation, the formulae of the differential decay rates of the processes $B_{s}(B^{*}_{s})\rightarrow D^{*}_{s1}(D^{'}_{s2})\ell\overline{\nu}$ appear as
\begin{align}
\label{rate1}
\frac{d\Gamma}{dy}(B_{s}\rightarrow D^{*}_{s1}\ell\overline{\nu})= &
\frac{G^{2}_{F}|V_{cb}|^2 m^{2}_{B_{s}}m^{3}_{D^{*}_{1s}}}{72\pi^3} |\xi(y)|^2 (y-1)^{5/2} (y+1)^{3/2} [(r_{1}^2+1) (2 y+1)\nonumber \\&-2 r_{1} \left(y^2+y+1\right)],
\\\label{rate2}
\frac{d\Gamma}{dy}(B_{s}\rightarrow D^{'}_{s2}\ell\overline{\nu})= &
\frac{G^{2}_{F}|V_{cb}|^2 m^{2}_{B_{s}} m^{3}_{D^{'}_{2s}}} {72\pi^3} |\xi(y)|^2 (y-1)^{5/2} (y+1)^{3/2} [(r_{2}^2+1) (4 y-1)\nonumber \\& -2 r_{2} \left(3 y^2- y+1\right)],
\\\label{rate3}
\frac{d\Gamma}{dy}(B^{*}_{s}\rightarrow D^{*}_{s1}\ell\overline{\nu})= &
\frac{G^{2}_{F}|V_{cb}|^2 m^{2}_{B^{*}_{s}}m^{3}_{D^{*}_{1s}}}{216\pi^3} |\xi(y)|^2 (y-1)^{5/2} (y+1)^{3/2} [(r_{3}^2+1) (7 y-1)\nonumber \\& -2 r_{3} \left(5 y^2-y+2\right)],
\\\label{rate4}
\frac{d\Gamma}{dy}(B^{*}_{s}\rightarrow D^{'}_{s2}\ell\overline{\nu})= &
\frac{G^{2}_{F}|V_{cb}|^2 m^{2}_{B^{*}_{s}} m^{3}_{D^{'}_{2s}}} {216\pi^3} |\xi(y)|^2 (y-1)^{5/2} (y+1)^{3/2} [(r_{4}^2+1) (11 y+1)\nonumber \\& -2 r \left(7 y^2+y+4\right)],
\end{align}
while for the processes $B_{s}(B^{*}_{s})\rightarrow D_{s2}(D^{*}_{s3})\ell\overline{\nu}$, they can be found to be
\begin{align}
\label{rate5}
\frac{d\Gamma}{dy}(B_{s}\rightarrow D_{s2}\ell\overline{\nu})= &
\frac{G^{2}_{F}|V_{cb}|^2 m^{2}_{B_{s}}m^{3}_{D_{2s}}} {1000\pi^3} |\zeta(y)|^2 (y-1)^{5/2} (y+1)^{7/2} [(r_{5}^2+1) (7 y-3) \nonumber \\& -2r_{5} \left(4 y^2 -3 y+3\right)],
\\\label{rate6}
\frac{d\Gamma}{dy}(B_{s}\rightarrow D^{*}_{s3}\ell\overline{\nu})= &
\frac{G^{2}_{F}|V_{cb}|^2m^{2}_{B_{s}}m^{3}_{D^{*}_{3s}}} {360\pi^3} |\zeta(y)|^2 (y-1)^{5/2}(y+1)^{7/2}[(r_{6}^2+1) (11 y+3)\nonumber \\& -2 r_{6} \left(8 y^2+3 y+3\right)],
\\\label{rate7}
\frac{d\Gamma}{dy}(B^{*}_{s}\rightarrow D_{s2}\ell\overline{\nu})= &
\frac{G^{2}_{F}|V_{cb}|^2 m^{2}_{B^{*}_{s}}m^{3}_{D_{2s}}} {3000\pi^3} |\zeta(y)|^2 (y-1)^{5/2} (y+1)^{7/2} [(r_{7}^2+1) (23 y+3)\nonumber \\& -2 r_{7} \left(16 y^2+3 y+7\right)],
\\\label{rate8}
\frac{d\Gamma}{dy}(B^{*}_{s}\rightarrow D^{*}_{s3}\ell\overline{\nu})= &
\frac{G^{2}_{F}|V_{cb}|^2 m^{2}_{B^{*}_{s}}m^{3}_{D^{*}_{3s}}} {1080\pi^3} |\zeta(y)|^2 (y-1)^{5/2}(y+1)^{7/2}[(r_{8}^2+1) (31 y-3)\nonumber \\& -2r_{8} \left(20 y^2-3 y+11\right)],
\end{align}
where $r_{i}$ ($i=1, \cdots, 8$) is the ratio between the mass of the final $\bar{c}s$ meson and that of the initial $\bar{b}s$ meson in each process, e.g., $r_{1}=\frac{M_{D^{*}_{s1}}}{M_{B_{s}}}$. The maximal values of $y$ for these semileptonic processes are given in Table \ref{table1}.
\begin{table}[h]
\caption{The maximal value of $y$ for each process: $y_{\text{max}}=(1+r_{i}^{2})/2r_{i}$ ($i=1, 2, \cdots, 8$).}
\begin{center}
\begin{tabular}{ccccccccc}
\hline \hline
 & & $D^{*}_{s1}\ell\overline{\nu}$ & & $D'_{s2}\ell\overline{\nu}$ & & $D_{s2}\ell\overline{\nu}$ & & $D^{*}_{s3}\ell\overline{\nu}$\\
\hline
$B_{s}$  & & 1.20493 & & 1.21673 & & 1.21427 & & 1.20457 \\
$B^{*}_{s}$ & & 1.21105 & & 1.22304 & & 1.22055 & & 1.21069 \\
\hline \hline
\end{tabular}
\end{center}\label{table1}
\end{table}
In addition, we need the input parameters $V_{cb}=0.04$ and $G_{F}=1.166\times10^{-5}\mbox{GeV}^{-2}$. By integrating the differential decay rates over the kinematic region $1.0 \leq y \leq y_{\text{max}}$, we get the decay widths of these semileptonic decay modes which are listed in Table \ref{table2}.
\begin{table}[h]
\caption{Predictions for the decay widths and branching ratios }
\begin{center}
\begin{tabular}{ccccc}
  \hline
  \hline
  Decay mode & & Decay width (GeV)  & & Branching ratio  \\
  \hline
  $B^{0}_{s}\rightarrow D^{*}_{s1}\ell\overline{\nu}$ & & $1.25^{+0.80}_{-0.60}\times10^{-19}$ & & $2.85^{+1.82}_{-1.36}\times 10^{-7}$ \\
  $B^{0}_{s}\rightarrow D^{'}_{s2}\ell\overline{\nu}$ & & $1.49^{+0.97}_{-0.73}\times10^{-19}$ & & $3.40^{+2.21}_{-1.66}\times 10^{-7}$ \\
  $B^{*}_{s}\rightarrow D^{*}_{s1}\ell\overline{\nu}$ & & $0.96^{+0.62}_{-0.46}\times10^{-19}$ & & $1.38^{+0.88}_{-0.67}\times 10^{-12}$ \\
  $B^{*}_{s}\rightarrow D^{'}_{s2}\ell\overline{\nu}$ & & $2.19^{+1.46}_{-1.08}\times10^{-19}$ & & $3.13^{+2.08}_{-1.54}\times 10^{-12}$ \\
  \hline
  $B^{0}_{s}\rightarrow D_{s2}\ell\overline{\nu}$ & & $4.48^{+1.05}_{-0.94}\times10^{-17}$ & & $1.02^{+0.24}_{-0.21}\times 10^{-4}$ \\
  $B^{0}_{s}\rightarrow D^{*}_{s3}\ell\overline{\nu}$ & & $1.52^{+0.35}_{-0.31}\times10^{-16}$ & & $3.46^{+0.80}_{-0.70}\times 10^{-4}$ \\
  $B^{*}_{s}\rightarrow D_{s2}\ell\overline{\nu}$ & & $5.12^{+1.20}_{-1.07}\times10^{-17}$ & & $7.31^{+1.72}_{-1.52}\times 10^{-10}$ \\
  $B^{*}_{s}\rightarrow D^{*}_{s3}\ell\overline{\nu}$ & & $1.74^{+0.40}_{-0.36}\times10^{-16}$ & & $2.49^{+0.57}_{-0.52}\times 10^{-9}$ \\
  \hline
  \hline
\end{tabular}
\end{center}\label{table2}
\end{table}
Notice that the lifetime of $B^{0}_{s}$ meson is $\tau_{B^{0}_{s}} = 1.5 ps$ , which means the total decay width is about $\Gamma_{B^{0}_{s}} = 4.388\times 10^{-13} \text{GeV}$. There has been no experimental result for the total width of the $B^{*}_{s}$ meson by now, but we know that its dominant decay mode is the radiative decay $B^{*}_{s}\rightarrow B_{s}\gamma$ \cite{PDG14},  the width of which is calculated theoretically to be about $\Gamma_{B^{*}_{s}} = 0.07 \text{keV}$ \cite{EFG02,Choi07}. We can take it as the total width of $B^{*}_{s}$ meson for a rough estimation for the branching rations of its semileptonic decays. Taking all these into account, we get the final branching ratios of the semileptonic decays mentioned above (see Table \ref{table2}). It is worth noting that the large errors in decay widths of $(B^{0}_{s}, B^{*}_{s})\rightarrow (D^{*}_{s1}, D^{'}_{s2})\ell\overline{\nu}$ are due to the relative large error in the form factor $\xi(y)$, which comes from the systematical uncertainty of the QCD sum rule approach. It can be expected that the $1/m_Q$ corrections may provide significant modification of the decay rates and improve the precision of the results, which may be taken into account in further works. As we can see in Table \ref{table2}, the branching ratio of $B^{*}_{s}$ semileptonic decays into the $D^{*}_{s1}(2860)$ and $D^{*}_{s3}(2860)$ are too small to be observed, while the branching ratios of $B^{0}_{s}$ semileptonic decays into these states are large enough to be measured by future experiments, such as the LHCb experiment.

In summary, we have studied the semileptonic decays of the ground state $\bar{b}s$ meson doublet $(0^-, 1^-)$ into the $1D$ excited family of $\bar{c}s$ meson, including the newly observed $D^{*}_{s1}(2860)$ and $D^{*}_{s3}(2860)$ mesons by the LHCb collaboration. In the framework of HQET, we have employed the QCD sum rule approach to estimate the leading-order universal form factors describing these weak transitions. With these universal form factors, the decay widths and branching ratios are estimated. We find that the decay widths are $\Gamma(B_s\rightarrow D^{*}_{s1}\ell\overline{\nu}) =1.25^{+0.80}_{-0.60}\times10^{-19} \mbox{GeV}$, $\Gamma(B_s\rightarrow D^{'}_{s2}\ell\overline{\nu}) =1.49^{+0.97}_{-0.73}\times10^{-19} \mbox{GeV}$, $\Gamma(B_s\rightarrow D_{s2}\ell\overline{\nu}) =4.48^{+1.05}_{-0.94}\times10^{-17} \mbox{GeV}$, and $\Gamma(B_s\rightarrow D^{*}_{s3}\ell\overline{\nu}) = 1.52^{+0.35}_{-0.31}\times10^{-16} \mbox{GeV}$. The corresponding branching ratios are $\mathcal {B}(B_s\rightarrow D^{*}_{s1}\ell\overline{\nu}) =2.85^{+1.82}_{-1.36}\times 10^{-7}$, $\mathcal {B}(B_s\rightarrow D^{'}_{s2}\ell\overline{\nu}) =3.40^{+2.21}_{-1.66}\times 10^{-7}$, $\mathcal {B}(B_{s}\rightarrow D_{s2}\ell\overline{\nu}) =1.02^{+0.24}_{-0.21}\times 10^{-4}$, and $\mathcal {B}(B_s\rightarrow D^{*}_{s3}\ell\overline{\nu}) = 3.46^{+0.80}_{-0.70}\times 10^{-4}$. We find that the branching ratios of some processes are large enough to be observed in future experiments. Measurements of these processes will be helpful for clarifying the properties of the orbitally $1D$ excited family of $\bar{c}s$ meson, such as mixing in these states.

\begin{acknowledgments}
This work was supported in part by the National Natural Science Foundation of China (NNSFC) under Grants Nos. 11205242, 11275268, 11475260, 11475259, 11305264, 11475258, 11405269, and the Reasearch Project of NUDT.
\end{acknowledgments}


\begin{thebibliography}{s2}
\bibitem{LHCb141}R. Aaij \textit{et al.} [LHCb Collaboration], Phys. Rev. Lett. \textbf{113}, 162001 (2014).
\bibitem{LHCb142}R. Aaij \textit{et al.} [LHCb Collaboration], Phys. Rev. D \textbf{90}, 072003 (2014).
\bibitem{BABAR06}B. Aubert \textit{et al.} [BaBar Collaboration], Phys. Rev. Lett. \textbf{97}, 222001 (2006).
\bibitem{BABAR09}B. Aubert \textit{et al.} [BaBar Collaboration], Phys. Rev. D \textbf{80}, 092003 (2009).
\bibitem{SCLM14}Q. T. Song, D. Y. Chen, X. Liu, and T. Matsuki, arXiv: 1408.0471 [hep-ph].
\bibitem{ZCCG14}D. Zhou, E. L. Cui, H. X. Chen, L. S. Geng, X. Liu, and S. L. Zhu, Phys. Rev. D \textbf{90}, 114035 (2014).
\bibitem{Wang14}Z. G. Wang, arXiv: 1408.6465 [hep-ph].
\bibitem{GM14}S. Godfrey and K. Moats, Phys. Rev. D \textbf{90}, 117501 (2014).
\bibitem{KZL14}H. W. Ke, J. H. Zhou, and X. Q. Li, arXiv: 1411.0376 [hep-ph].
\bibitem{D009}V. M. Abazov \textit{et al.} [D0 Collaboration], Phys. Rev. Lett. \textbf{102}, 051801 (2009).
\bibitem{ASS14}K. Azizi, H. Sundu, and S. Sahin, arXiv: 1411.3100 [hep-ph].
\bibitem{Neu92}M. Neubert, Phys. Rev. D \textbf{45}, 2451 (1992); \textbf{46}, 3914 (1992).
\bibitem{LLSW97}A. K. Leibovich, Z. Ligeti, I. W. Stewart, and M. B. Wise, Phys. Rev. Lett. \textbf{78}, 3995 (1997); Phys. Rev. D \textbf{57}, 308 (1998).
\bibitem{Dea98}A. Deandrea, N. Di Bartolomeo, R. Gatto, G. Nardulli, and A. D. Polosa, Phys. Rev. D \textbf{58},
034004 (1998); V. Mor\'{e}nas, A. Le Yaouanc, L. Oliver, O. P\`{e}ne, and J. C. Raynal, Phys. Rev. D \textbf{56}, 5668 (1997).
\bibitem{EFG99}D. Ebert, R. N. Faustov, and V. O. Galkin, Phys. Rev. D \textbf{61}, 014016 (1999); \textbf{75}, 074008 (2007).
\bibitem{Hua99}M. Q. Huang and Y. B. Dai, Phys. Rev. D \textbf{59}, 034018 (1999); \textbf{64}, 014034 (2001).
\bibitem{Col00}P. Colangelo, F. De Fazio, and G. Nardulli, Phys. Lett. B \textbf{478}, 408 (2000).
\bibitem{Hua04}M. Q. Huang, Phys. Rev. D \textbf{69}, 114015 (2004).
\bibitem{AAO06}T. M. Aliev, K. Azizi, and A. Ozpineci, Eur. Phys. J. C \textbf{51}, 593 (2007); T. M. Aliev and M. Savci, Phys. Rev. D \textbf{73}, 114010 (2006).
\bibitem{Gan09}L. F. Gan and M. Q. Huang, Phys. Rev. D \textbf{79}, 034025 (2009); \textbf{79}, 117501 (2009).
\bibitem{SAEFHP11}J. Segovia, C. Albertus, D. R. Entem, F. Fernandez, E. Hernandez, and M. A. Perez-Garcia, Phys. Rev. D \textbf{84}, 094029 (2011).
\bibitem{Shi79}M. A. Shifman, A. I. Vainshtein, and V. I. Zakharov, Nucl. Phys. \textbf{B147}, 385 (1979); \textbf{B147}, 448 (1979); V. A. Novikov, M. A. Shifman, A. I. Vainshtein, and V. I. Zakharov, Fortschr. Phys. \textbf{32}, 11 (1984).
\bibitem{Neu94}M. Neubert, Phys. Rep. \textbf{245}, 259 (1994), and references therein; Aneesh V. Manohar and Mark B. Wise, \textsl{Heavy Quark Physics} ( Cambridge University Press, New York, 2000).
\bibitem{Gan10}L. F. Gan and M. Q. Huang, Phys. Rev. D \textbf{82}, 054035 (2010); Commun. Theor. Phys. \textbf{57}, 85 (2012).
\bibitem{Fal92}A. F. Falk, Nul. Phys. \textbf{B378}, 79 (1992); A. F. Falk and M. Luke, Phys. Lett. B \textbf{292}, 119 (1992).
\bibitem{FeynC}http://www. feyncalc.org/.
\bibitem{Dai97}Y. B. Dai, C. S. Huang, M. Q. Huang, and C. Liu, Phys. Lett. B \textbf{390}, 350 (1997); Y. B. Dai, C. S. Huang, and M. Q. Huang, Phys. Rev. D \textbf{55}, 5719 (1997).
\bibitem{Blo93}B. Blok and M. Shifman, Phys. Rev. D \textbf{47}, 2949 (1993).
\bibitem{PDG14}K. A. Olive \textit{et al.} (Particle Data Group), Chin. Phys. C \textbf{38}, 090001 (2014).
\bibitem{DHLZ03}Y. B. Dai, C. S. Huang, C. Liu, and S. L. Zhu, Phys. Rev. D \textbf{68}, 114011 (2003).
\bibitem{EFG02}D. Ebert, R. N. Faustov, and V. O. Galkin, Phys. Lett. B \textbf{537}, 241 (2002).
\bibitem{Choi07}H. M. Choi, Phys. Rev. D \textbf{75}, 073016 (2007).

\end{thebibliography}
\end{document}